\documentstyle[12pt]{article}

\title{\bf Renormalised perturbative series\\
           for the impurity levels\\
           in a quantum well}

\author{Augusto Gonzalez$^{*,**}$, Ilia Mikhailov$^{***}$}

\date{$^*$Depto. de Fisica, Univ. Nacional de Colombia, Sede Medellin,\\
       AA 3840, Medellin, Colombia\\
      $^{**}$Instituto de Cibernetica, Matematica y Fisica, Calle E 309,\\
       Vedado, Habana 4, Cuba\\
      $^{***}$Escuela de Fisica, Universidad Industrial de Santander,\\
       AA 568, Bucaramanga, Colombia}

\begin{document}

\maketitle

\begin{abstract}

The energy levels of an impurity center in a deep quantum well of width
$L$ and depth $g$ are studied analytically . Renormalised perturbative
series are constructed in the regions $g L^2 << 1$ and $g L^2 >> 1$.
Maximal binding energy and wave function deformation to a
quasi-twodimensional function are found to occur at a certain $L_c$
satisfying $\sqrt{g} L_c \sim 1$. Similar results may be obtained for
the impurity in a quantum wire, in a dot or in a multiwell structure.

\end{abstract}
\newpage

\section{Introduction}

The energy levels of donor and acceptor centers in quantum well, wires
and dots have been widely studied recently in the literature [1].

In the present paper, we show that very simple analytic estimates to the
energy levels may be obtained from perturbation theory. The perturbative
series are renormalised to account for large orders corrections or, said
in a different way, the renormalised series are forced to interpolate
between the different expansions obtained in different regimes.

In a deep quantum well of deepth $g$ and width $L$ (in effective atomic
units), we construct renormalised series for $g L^2 << 1$, and $g L^2 >>
1$. The starting points are the perturbative expansions obtained in the
following limits:

\begin{description}
\item{i)} $g L^2 << 1$, $g L << 1$. Shallow level, three dimensional
          coulombic wave function.
\item{ii)} $g L^2 << 1$, $g L >> 1$. Shallow level, quasi two-dimensional
           wave function out of the well.
\item{iii)} $g L^2 >> 1$, $L << 1$. Deep level, quasi two-dimensional
           wave function inside the well.
\item{iv)} $g L^2 >> 1$, $L >> 1$. Deep level, three dimensional coulombic
           wave function.
\end{description}

At high values of $g$, the curve $E_b(L)$, i.e. the binding energy as a
function of $L$, contains the four regimes i) -- iv). Indeed, let us
consider, for example, $g = 60$. Then, when $L < 1/60$ the conditions for
i) are fulfilled. In the interval $1/60<L<1/\sqrt{60}$, we may use an
approximation like ii). When $1/\sqrt{60}<L<1$, the regime iii) holds and,
finally, at $L > 1$, the wave function is certainly as indicated in iv).

The transition from ii) to iii) is found to ocur at $\sqrt{g} L \sim 1$, and
corresponds to a maximum of the binding energy.

\section{Impurity in a quantum well}

We start from the effective mass hamiltonian in effective atomic units

\begin{equation}
H = -\frac{1}{2} \Delta - \frac{1}{r} + g \Theta(z),
\end{equation}

\noindent
where the energy unit is $m e^4/(\hbar^2\kappa^2)$, $m$ is the electron
effective mass, $\kappa$-- the relative  dielectric constant, $g=V_0/[m e^4/
(\hbar^2 \kappa^2)]$ is the well depth, and $\Theta$ is the step function

\begin{equation}
\Theta(z) = \left\{
\begin{array}{ll}
0, & z_1<z<z_2, ~~z_2-z_1=L,\\
1, &{\rm outside}.
\end{array}
\right.
\end{equation}

Typically, $g\ge 20$. From the eigenvalue of $H$, $E$, the binding energy
is defined as

\begin{equation}
E_b = E_w - E,
\end{equation}

\noindent
where $E_w$ is the threshold for the continuous spectrum in the well, i.e.
the lowest eigenvalue of the one-dimensional hamiltonian

\begin{equation}
H_w = -\frac{1}{2} {{\rm d}^2 \over {\rm d}z^2} + g \Theta(z).
\end{equation}

\noindent
It is found from $E_w = k^2/2$, where $k$ is the smallest solution of

\begin{equation}
\cos \left({k L\over 2}\right) = \sqrt{2\over g L^2} \left({k L \over 2}
                                 \right).
\end{equation}

The asymptotic expressions for $E_w$ are the following

\begin{eqnarray}
E_w &=& g \left\{1 - \frac{1}{2} g L^2 + \frac{1}{3} g^2 L^4 +\dots
        \right\}, ~~~g L^2 << 1\\
E_w &=& {\pi^2 \over L^2} \left\{\frac{1}{2} - \sqrt{{2\over g L^2}} +
        {3\over g L^2} - \left(2+{\pi^2\over 24}\right)
        \left({2\over g L^2}\right)^{3/2} + \dots \right\}, \nonumber\\
    & & g L^2 >> 1.
\end{eqnarray}

We shall obtain analytic estimates to $E$ to compute the binding energy
as a function of $L$. The parameter $g$ is assumed to be large.

The exact quantum number of the present problem is the $z$-projection of
angular momentum, which will be called $J$. Without loss of generality, we
may consider only positive values of $J$. To avoid using degenerate
perturbation
theory, we will study the first level with a given $J$. when $L\to 0$
or $L\to\infty$, we recover the unbounded (3D) coulomb problem, and the
states we study have angular momentum equal to $J$.

\section{The renormalised series for $g L^2 << 1$}

When $g L^2 << 1$, the levels are shallow, i.e. located near the top of
the barrier, and the wave function is mainly outside the well. The
characteristic confinement distance along the $z$-direction is $l_z$, where
$l_z^{-1}=\sqrt{2(g-E_w)}=g L+\dots$. When $g L << 1$, the wave function is
basically the 3D coulombic wave function, $\phi_{3D}$. On the other hand,
when $g L >> 1$ (keeping $g L^2 << 1$, i.e. $l_z >> L$) the confinement
length is much less than the Bohr radius (which in our units is one), and
the wave function is written approximately as $\exp{-z/l_z}~\phi_{2D}$,
where $\phi_{2D}$ is the 2D coulombic wave function.

Accordingly, when $g L << 1$ we write

\begin{equation}
H = H_{3D} + g + V_1,
\end{equation}

\noindent
where $V_1=-g \Theta^*(z)$, and the complementary step function is defined
from $\Theta(z)+\Theta^*(z)=1$. $V_1$ will be treated as perturbation.
From first order perturbation theory, we obtain

\begin{eqnarray}
E &=& g - \frac{1}{2(J+1)^2} - \frac{g L}{2^{J+1} J! (J+1)^{J+1}}
          \nonumber \\
  &+& \frac{g L^3 (\varsigma_2^2+\varsigma_1\varsigma_2+\varsigma_1^2)}
          {J! (J+1)^{J+4} 2^J} + {\cal O}(L^5),
\end{eqnarray}

\noindent
where we have written, $z_1=\varsigma_1 L$, $z_2=\varsigma_2 L$,
$\varsigma_2-\varsigma_1=1$, and without loss of generality, we will
assume that $\varsigma_1 < 0$, $\varsigma_2 > 0$. The leading
contribution of second order perturbation theory will be $-0.002~g^2 L^2$
for the ground state ($J=0$), and practically zero for $J>0$.

Thus, groupping (6) and (9), we obtain for the binding energy at $g L << 1$,

\begin{equation}
E_b = a_0 + a_1 (g L) + a_2 (g L)^2 + a_3 (g L)^3 + \dots,
\end{equation}

\noindent
where $a_0=1/(2 (J+1)^2)$, $a_1=1/(2^{J+1} J! (J+1)^{J+1})$, $a_2=-0.498$
when $J=0$ and $a_2=-1/2$ when $J\ge1$, $a_3=-(\varsigma_2^2+
\varsigma_1\varsigma_2+\varsigma_1^2)/(g^2 J! (J+1)^{J+4} 2^J)$, etc.

Now, let us consider the opposite limit, $g L >> 1$. The hamiltonian will
be written as

\begin{equation}
 H= H_{2D} + H_w + V_2,
\end{equation}

\noindent
where $V_2=1/\rho-1/\sqrt{\rho^2+z^2}$ will be considered as perturbation.
$\rho$ is the polar coordinate in the plane. In first order perturbation
theory, we obtain

\begin{equation}
E = -\frac{1}{2 (J+1/2)^2} + E_w -
     \frac{b_k}{(g L)^k} + \dots,
\end{equation}

\noindent
where the first correction is $b_1=-8$ for the g.s. ($J=0$), and $b_2=
-2 (2 J-2)!/(2 J+1)!/(J+1/2)^3$ for the excited states ($J\ge1$).
Consequently, for the binding energy, we obtain

\begin{equation}
E_b = \frac{b_k}{(g L)^k} + b_0 + \dots,
\end{equation}

\noindent
where $b_0=(J+1/2)^{-2}/2$. Notice that Eq. (13) may be applied to the g.s.
when $g L\ge 8$. If we recall that $g L^2 << 1$, then $g$ is forced to be
greater than 64.

Note that the first three terms of (10) and the first two terms of (13)
show that $E_b$ is, in certain approximation, ``universal'', in the
sense that it depends only on the variable $g L$. If we consider corrections
such as the term $a_3 (g L)^3$ of (10), this universality is lost. Notice
also that the impurity position appears for the first time precisely in
$a_3$.

Once we have the correct behaviour of $E_b$ for large values of $g L$, we
may construct a ``renormalised'' series from (10) such that the large
orders of this series will account for the correct asymptotics at
$g L >> 1$. We follow the idea of paper [2] in which the method was applied
to the two-electron problem in a quantum dot.

For simplicity, we consider the g.s. of the centered impurity and include
the first two terms of (10). We write $\beta=g L/(\alpha+g L)$. When
$g L\to 0$, $\beta\to 0$, whereas when $g L \to\infty$, $\beta\to 1$.
$\alpha$ is a free parameter which will be used to fit the numerical
results. It gives an idea of where the transition from ``small'' $g L$ to
``large'' $g L$ takes place. The renormalised series is looked for as a
series in $\beta$

\begin{equation}
E_b = c_0 + c_1 \beta + c_2 \beta^2 + c_3 \beta^3 + \dots.
\end{equation}

The first coeficients $c_k$ are defined in such a way that when $\beta\to 0$
the first terms of (10) are reproduced. That is: $c_0=1/2$, $c_1=\alpha/2$.
On the other hand, $c_2$ and $c_3$ are chosen to reproduce
the correct behaviour at $\beta\to 1$. We note that, as $\beta\to 1$

\begin{equation}
E_b = (c_0+c_1+c_2+c_3)-(c_1+2 c_2+3 c_3)(1-\beta) + \dots,
\end{equation}

\noindent
that is, $c_3$ and $c_4$ are required to satisfy: $1/2+\alpha/2+c_2+c_3=2$,
$\alpha/2+2 c_2+3 c_3=8/\alpha$.

We show in Fig. 1 variational computations (points) for the g.s. of the
centered impurity at $g=60, ~100, ~140$ and 180. At each $g$, values of
$L$ observing $L \le 1/\sqrt{g}$ were included. A trial function with two
nonlinear parameters was used in the computations [3]. The results show a
very small dispersion of the points around a fixed curve, in accordance
with the predicted universality. The curve is well fitted by the series
(14) with $\alpha \approx 4$ (the solid line).

\section{The renormalised series for $g L^2 >> 1$}

Now, we consider the situation in which the level is deep inside the well,
$g L^2 >> 1$. At large values of $g$, we have again two limiting situations:
$L<<1=a_B$, and $L>>1$. The corresponding wave functions are approximately
$\phi_{2D}~\sin[\pi (z-z_1)/L]$, and $\phi_{3D}$.

When $L>>1$, we take $V_3=g\Theta(z)$ as a perturbation to $H_{3D}$.
Assuming that $z_1$ and $z_2$ are both finite, we obtain

\begin{equation}
E = -{1\over 2 (J+1)^2} + {\cal O}\left(e^{-\varsigma L}\right),
\end{equation}

\noindent
where $\varsigma$ is the minimum between $|\varsigma_1|$ and $\varsigma_2$.
The binding energy is thus

\begin{equation}
E_b = b'_0 + {b'_2\over L^2} + {b'_3\over L^3} + \dots,
\end{equation}

\noindent
where $b'_0=(J+1)^{-2}/2$, $b'_2=\pi^2/2$, $b'_3=-\pi^2\sqrt{2/g}$, etc.

On the other hand, when $L<<1$ we write an expression like (11), and
consider $V_2$ as a perturbation. To first order, we get

\begin{equation}
E = -{1\over 2 (J+1/2)^2} + E_w - a'_k L^k + \dots,
\end{equation}

\noindent
where the first correction is $a'_1=-16[1/2+\varsigma_1+\varsigma_1^2-
(1-\cos{2\pi\varsigma_1})/(2\pi^2)]$ for the g.s. ($J=0$), and $a'_2=
-4 (2 J-2)! (\varsigma_1^2+\varsigma_1+1/3-1/2/\pi^2)/(2 J+1)!/(J+1/2)^3$
for the excited states ($J\ge 1$). That is,

\begin{equation}
E_b = a'_0 + a'_k L^k + \dots,
\end{equation}

\noindent
where $a'_0=1/2/(J+1/2)^2$.

Notice that the series (17) and (19) suggest, again, a universal behaviour
of $E_b$ in the leading approximation. That is, $E_b$ will not depend on
$g$ when $g L^2 >> 1$.

For simplicity, we consider again the g.s. of the centered impurity and
include terms up to $1/L^2$ in (17). To construct the renormalised series
for $g L^2 >>1$, we write $\beta'=L/(\alpha'+L)$,

\begin{equation}
E_b = c'_0+c'_1 (1-\beta')+c'_2 (1-\beta')^2+c'_3 (1-\beta')^3
    + c'_4 (1-\beta')^4+\dots.
\end{equation}

The coefficients $c'_0$, $c'_1$ and  $c'_2$ are obtained from (17), i.e.
$c'_0=1/2$, $c'_1=0$, $c'_2=\pi^2/2/\alpha'^2$. The coefficients $c'_3$
and $c'_4$ are required to satisfy the linear equations
$1/2+\pi^2/2/\alpha'^2+c'_3+c'_4=2$, $\pi^2/\alpha'^2+3 c'_3+4 c'_4=
\alpha'(4-16/\pi^2)$.

We show in Fig. 2 how close the behaviour of $E_b(L)$ is
to the universal behaviour when $g=60, ~100, ~140$ and 180. Variational
computations corresponding to values of $L$ for which $L \ge 2.6/\sqrt{g}$
are represented as points. The dispersion of the points is very small. The
universal curve is fitted well by the series (20) when the parameter
$\alpha'$ is near to 1. The result of the fit is presented as a solid line.

\section{The maximum of the curve $E_b$ vs $L$}

The conclusion of the previous analysis is the following. As $L$ is
increased from 0 to $1/\sqrt{g}$ the wave function changes from a
tridimensional coulombic wave function to a quasi-twodimensional function
out of the well. The binding energy is a universal function of $g L$. On
the other hand, as $L$ is decreased from infinity to $2.6/\sqrt{g}$ the
wave function undergoes a change from the tridimensional coulombic wave
function to a quasi-twodimensional function inside the well. The binding
energy turns out to be a universal function of $L$. As a result of
compressing the wave function, the binding energy increases. There is a
critical value, $L_c$, at which the wave function is maximally compressed
and $E_b$ is maximal. As indicated,

\begin{equation}
{1 \over \sqrt{g}} < L_c < {2.6 \over \sqrt{g}}.
\end{equation}

We may obtain a rough estimate to $L_c$ as the point at which the two
series (13) and (19) coincide. For the g.s. of the centered impurity, we get

\begin{equation}
L_c \approx {1.83 \over \sqrt{g}}.
\end{equation}

Numerical computations at $g \ge 60$ show that the product $\sqrt{g} L_c$
is almost constant, taking a value near 1.5 (Fig. 3).

We note that the dependence $L_c \sim 1/\sqrt{g}$ comes also from very
simple reasonings based on the indeterminacy relations [4].

\section{Concluding Remarks}

We have shown that, at large values of $g$, the parameters $g L$ and $g L^2$
identify different regimes in the behaviour of the binding energy and the
wave function of the impurity problem in a quantum well. In particular, the
binding energy was shown to be a ``universal'' function in each of the
regions $g L^2 << 1$ and $g L^2 >> 1$.

Our description is qualitatively valid in problems where $g=20~-~60$.
As $g$ is still decreased, regimes ii) and iii) become less applicable, and
may be absent at all . In particular, at $g\approx 1$, we expect a
transition directly from i) to iv) as $L$ goes from $L<<1$ to $L>>1$.

We would like to stress that there are similarities between the results of
our section 4 and paper [5], in which the $g \to \infty$ limit is studied.
In that paper, a decomposition like (11) is choosen in the entire interval
$0 < L < \infty$, and perturbation theory is applied. The failure of this
decomposition at $L \ge 1$ is
corrected by introducing a scaling parameter and asking for the virial
theorem to hold. Our approach is more qualitative but, at the same time,
more exact, stressing what the actual wave function really is at any $L$.

A qualitative analysis of wave functions and energy curves for the impurity
problem in dots, wires and multiwell structures is as simple as the analysis
presented above. It may be a useful complement to the existing (or in
progress) sophisticated numerical calculations.

\section{Acknowledgments}

One of the authors (A. G.) acknowledges financial support from the Colombian
Institute for Science and Technology (COLCIENCIAS) under Project
1118-05-661-95, and from the Committee for Scientific Research
at the Universidad Nacional (CINDEC). The authors are endebted to B.
Rodriguez, J. Marin and J. Betancur for useful discussions.

\newpage

{\Large Figure Captions}
\vspace{1cm}

\begin{description}
\item{Fig. 1.} Twice the binding energy of the centered impurity as a
               function of $g L$. Points are variational calculations
               satisfying $L \le 1/\sqrt{g}$.
\item{Fig. 2.} Twice the binding energy as a function of $L$ when $L \ge
               2.6/\sqrt{g}$.
\item{Fig. 3.} The dependence of $\sqrt{g} L_c$ on $g$.
\end{description}

\end{document}